\documentclass[12pt, a4paper, twocolumn]{article}

\usepackage{amssymb}
\usepackage{amsthm}
\usepackage[fleqn]{amsmath}

\usepackage{mathtools}

\usepackage{multirow}
\usepackage{hhline}
\usepackage{arydshln}

\usepackage{graphicx}
\usepackage{caption}
\usepackage{subcaption}
\usepackage{csquotes}
\usepackage[super]{nth}

\usepackage{authblk}

\newcommand{\argmax}{\arg\!\ \max} 
\DeclarePairedDelimiter\abs{\lvert}{\rvert}%
\DeclarePairedDelimiter\norm{\lVert}{\rVert}%
\DeclarePairedDelimiter{\ceil}{\lceil}{\rceil}

\title{Analysis of memory in LSTM-RNNs for source separation}

\author[1]{Jeroen Zegers}
\author[1]{Hugo Van hamme}
\affil[1]{ESAT, KU Leuven, Belgium}
\providecommand{\keywords}[1]
{
  \small	
  \textbf{\textit{Keywords---}} #1
}

\usepackage{natbib}

\begin{document}
\maketitle











\begin{abstract}
Long short-term memory recurrent neural networks (LSTM-RNNs) are considered state-of-the art in many speech processing tasks. The recurrence in the network, in principle, allows any input to be remembered for an indefinite time, a feature very useful for sequential data like speech. However, very little is known about which information is actually stored in the LSTM and for how long. We address this problem by using a memory reset approach which allows us to evaluate network performance depending on the allowed memory time span. We apply this approach to the task of multi-speaker source separation, but it can be used for any task using RNNs. We find a strong performance effect of short-term (shorter than 100 milliseconds) linguistic processes. Only speaker characteristics are kept in the memory for longer than 400 milliseconds. Furthermore, we confirm that performance-wise it is sufficient to implement longer memory in deeper layers. Finally, in a bidirectional model, the backward models contributes slightly more to the separation performance than the forward model.
\end{abstract}


\keywords{multi-speaker source separation, long short-term memory, recurrent neural networks, memory analysis}




\section{Introduction}
\label{sec:intro}
Deep learning has been dominant for many speech tasks in recent years \citep{deep-neural-networks-for-acoustic-modeling-in-speech-recognition}. Since speech is a dynamic process, sequential models like recurrent neural networks (RNNs) seem ideal for modeling the underlying process \citep{graves2013speech}. RNNs give state-of-the-art performance in many speech \citep{bahdanau_chorowski_serdyuk_brakel_bengio_2016, Yu2016permutation} and non-speech \citep{sundermeyer_ney_schluter_2015,NIPS2015_5955} related tasks. While standard RNN cells are theoretically capable to remember any input from the past and use this information as it sees fit, it is found that their memory time span (the duration for which information is kept in memory) is actually relatively short. This is caused by vanishing and exploding gradients, which occur when the recurrent network is unrolled through time for the backpropagation step \citep{bengio1994learning,mozer1992induction,hochreiter2001gradient}, a problem also observed in very deep networks. 

A solution to this gradient problem was given with the introduction of the long short-term memory (LSTM) cell \citep{hochreiter1997long}. The principle of the constant error carousel counters the vanishing and exploding gradients encountered with the regular RNN cell \citep{hochreiter1997long}. It was shown that these LSTM-RNNs succeeded in solving simple, artificial tasks with long range dependencies of over a thousand time steps. However, it is unknown how long the memory time span for LSTM cells is for real and complex tasks like speech processing.

The aim of this paper is to examine the time span and importance of internal dynamics in the LSTM memory. This will be done by resetting the state of the LSTM cell at particular time intervals as to limit the allowed memory time span. By gradually reducing the reset frequency, bigger memory time spans are allowed. The networks are evaluated for different reset frequencies and the task performance differences can be used to assess the importance of the different memory spans. Furthermore, it is possible to use different memory spans for different layers in the LSTM-RNN. 
This allows to confirm or reject the hypothesis that deeper layers in RNNs bring higher level abstractions of the data and therefore use a bigger time span \citep{chung2016hierarchical}. Finally, different memory spans can be used for the forward and backward direction of a bidirectional LSTM-RNN which allows to distinguish between the importance of the directions.

The task of multi-speaker source separation (MSSS) seems well suited for this analysis as it has been shown that both long-term and short-term effects are important \citep{zegers2018memory}. This paper therefore focuses on the MSSS task, but the proposed methodology can be applied to any task using RNNs. Specifically, we would like to answer the following research questions with regards to MSSS:
\begin{itemize}
    \item Which order of time spans are important when using an LSTM-RNN for MSSS and can we link these time spans with descriptions of speech like phonetics, phonotactics, lexicon, prosody and grammar?
    \item Since it is has been shown that speaker characterization is relevant for the task \citep{zegers2017improving,zegers2018memory}, can we find the amount of context necessary for the LSTM-RNN to sufficiently characterize the speakers in overlapping speech?
    \item For MSSS, do we observe the same hierarchical property that deeper layers have larger time dependencies, as was found by \cite{chung2016hierarchical}?
    \item In bidirectional LSTM-RNNs, would either direction be more important than the other for MSSS?
\end{itemize}

The rest of this paper is organized as follows. In section \ref{sec:lit} an overview of related work will be given and in section \ref{sec:MSSS} the task of MSSS will be explained, as well as how LSTM-RNNs can be used to tackle this problem. The memory reset LSTM cell is introduced in section \ref{sec:reset}. The experimental setup is given in section \ref{sec:exp} and results are discussed in section \ref{sec:res}. A final conclusion is given in section \ref{sec:conc}.

\section{Related work}
\label{sec:lit}
To our knowledge \cite{singh2016multi} is the only work where a similar reset approach to ours is given. In their paper a multi-stream system with an LSTM component for video action detection is described. A similar memory reset approach is used, but only for a unidirectional single layer LSTM. In this paper we extend to bidirectional multi-layer LSTMs where we allow layer dependent reset periods (section \ref{sec:reset_rnn}) and a method to reduce computational burden for longer reset periods (section \ref{sec:groups}). This makes the state reset algorithm far more complex compared to the unidirectional single layer reset. Furthermore, the memory reset by \cite{singh2016multi} was done during testing only, causing a train-test mismatch. In this work memory reset will be done during training and testing.

A different approach, called the segment approach in this paper for further reference, has a similar aim to restrict the LSTM memory span \citep{mohamed2015deep,46183,tuske2018investigation}. Segments are created by shifting a window by one time step over the input data. Each segment is passed through the LSTM and the output of the LSTM at the last time step within the segment is retained. The output of each segment has been produced with an LSTM with a memory span equal to the length of the window. It has been verified in our experiments that the reset approach and the segment approach give the same results. 

\cite{46183} and \cite{tuske2018investigation} used the segment approach on a language modeling task and claimed that perplexity scores did not improve by increasing the memory span over 40 time steps (words). Similarly, word error rates (WER) converged at 20 time steps. In both works the number of different memory lengths evaluated was rather limited and they were mainly interested in the performance saturation point. There was no analysis on the importance of different time scales within the model.

The segment approach was first applied by \cite{mohamed2015deep} on the spectral input of an automatic speech recognition (ASR) task. They found that the WER of the acoustic LSTM-RNN saturated relatively quickly. Therefore it was concluded that the main strength of the RNN is the frame-by-frame processing rather than the ability to have a large memory span. However, we found that for the task of MSSS long-term dependencies were, in fact, important for the separation quality.


A third approach to assess the relative importance of different memory span lengths is the leaky approach introduced by \cite{zegers2018memory}. There, a fraction of the LSTM cell state is leaked, on purpose, at every time step. This forces the LSTM to forget information of the past over time. The leaky approach can be seen as a soft reset compared to the hard reset of the reset approach proposed in this paper. The amount of computations in a leaky LSTM cell remains unchanged, while the computational load, for both the reset and segment approach, scale with the width of the memory span. The leaky approach is thus more interesting from a computational standpoint, but since the reset is soft, the timings found will not be exact.

The above approaches focus on analysis of the memory span by limiting memory capabilities. There are multiple works that give no explicit in-depth time analysis but adapt or restrict the memory span of the RNN in order to improve performance on a task. \cite{chung2016hierarchical} and \cite{el1996hierarchical} replaced the soft reset of the forget gate with a hard reset implementation. The network tries to learn the optimal reset frequency. Other approaches, like the clockwork RNN, have different, fixed update frequencies for different cells in the network. Memory restrictions are made on a cell level, rather than on a network or layer level. The idea is that some cells should focus on long-term effects and some should focus on short-term \citep{koutnik2014clockwork,alpay2016learning,neil2016phased}.





\section{Multi-speaker source separation}
\label{sec:MSSS}
In this section a well known MSSS algorithm called Deep Clustering (DC) \citep{hershey2016deep} will be explained. The method is invariant to permutations in the order of the reference speakers. It relies on intrinsic speaker characterization by the network to consistently map time-frequency bins dominated by the same speaker to the same point in the output space for a given mixture. At the end of the section, i-vectors, an explicit speaker characterization embedding often used in speaker recognition tasks, will be given as an alternative to this intrinsic speaker characterization. The i-vector of the active speakers can simply be appended to the input data. In the result section it will be shown that this increases the separation performance for short-term memory spans since the network is unburdened from the speaker characterization subtask.

\subsection{Task and permutation problem}
\label{sec:task}
When a mixture signal $y[n]=\sum_{s=1}^{S}x_s[n]$ of $S$ speakers is presented, the goal of MSSS is to estimate a signal $\hat{x}_s[n]$ for the $s^{th}$ speaker that is as close as possible to the source signal $x_s[n]$. This task can be expressed in the time-frequency domain using the short-time Fourier transform (STFT) of the signals. $\hat{X}_s(t,f)$ should then be estimated from $Y(t,f)=\sum_{s=1}^{S}X_s(t,f)$. The inverse STFT (ISTFT) can be used to find $\hat{x}_s[n]$ from $\hat{X}_s(t,f)$. Typically, for each speaker a mask $\hat{M}_s(t,f)$ is estimated such that 
\begin{equation}
\label{eq:spec_est}
\hat{X}_s(t,f)=\hat{M}_s(t,f)Y(t,f),
\end{equation}
for every time frame $t=0, \ldots, T-1$ and every frequency $f=0, \ldots, F-1$. One approach to address the MSSS task is to find mappings $g_{tf}$ from the input mixture to the mask estimates:
\begin{equation}
\label{eq:mask_est}
\hat{M}_s(t,f)=g_{tf}(\mathbf{Y}),
\end{equation}
with the constraint that $\hat{M}_s(t,f)\ge 0$ and $\sum_{s=1}^{S}\hat{M}_s(t,f)=1$ for every time-frequency bin $(t,f)$. In this paper $g_{tf}$ will be modeled with an LSTM-RNN. A differentiable loss function can be used to assess the quality of the speech estimates
\begin{equation}
\label{eq:general_loss}
\mathcal{L} = \sum_{s=1}^{S}\sum_{t,f} \mathcal{D}(\abs{\hat{X}_{s}(t,f)},\abs{X_s(t,f)}),
\end{equation}
with $\mathcal{D}$ some discrepancy measure. However, since an intra-class separation task is executed and no prior information on the speakers is assumed to be known, there is no guarantee that the network's assignment of speakers is consistent with the speaker labels of the targets. This is referred to as the label ambiguity or permutation problem \citep{hershey2016deep}. To cope with this ambiguity, a loss function has to be defined that is independent of the order of the target speakers. DC uses such a permutation invariant loss function.

\subsection{Deep Clustering}
\label{sec:DC}
In DC, a $D$-dimensional embedding vector $\mathbf{v}_{tf}$ is constructed for every time-frequency bin as $\mathbf{v}_{tf}=g_{tf}(\mathbf{Y})$, where $\mathbf{v}_{tf}$ has unit length. A ($TF\times D$)-dimensional matrix $\mathbf{V}$ is then constructed from these embedding vectors. Similarly, a ($TF\times S$)-dimensional target matrix $\mathbf{Z}$ is defined. If target speaker $s$ is the dominant speaker for bin $(t,f)$, then $z_{tf,s}=1$, otherwise $z_{tf,s}=0$. Speaker $s$ is dominant in a bin $(t,f)$ if $s=\argmax_{s'}(\abs{X_{s'}(t,f)})$. A permutation independent loss function is then defined as
\begin{equation}
\begin{split}
\label{eq:dc_loss}
\mathcal{L} &= \norm{\mathbf{V}\mathbf{V}^T-\mathbf{Z}\mathbf{Z}^T}_F^2 \\
&= \sum_{t_1,f_1,t_2,f_2}(\langle \mathbf{v}_{t_1f_1},\mathbf{v}_{t_2f_2}\rangle-\langle \mathbf{z}_{t_1f_1},\mathbf{z}_{t_2f_2}\rangle)^2,
\end{split}
\end{equation}
where $\norm{.}_F^2$ is the squared Frobenius norm. Since $\mathbf{z}_{tf}$ is a one-hot vector,
\begin{equation}
\langle \mathbf{z}_{t_1f_1},\mathbf{z}_{t_2f_2}\rangle=
\begin{cases}
      1, & \text{if}\ \mathbf{z}_{t_1f_1}=\mathbf{z}_{t_2f_2} \\
      0, & \text{otherwise}
    \end{cases}.
\end{equation}
The ideal angle $\phi_{t_1f_1,t_2f_2}$ between the normalized vectors $\mathbf{v}_{t_1f_1}$ and $\mathbf{v}_{t_2f_2}$ is thus
\begin{equation}
\phi_{t_1f_1,t_2f_2}=
\begin{cases}
      0, & \text{if}\ \mathbf{z}_{t_1f_1}=\mathbf{z}_{t_2f_2} \\
      \pi/2, & \text{otherwise}
    \end{cases}.
\end{equation}
After estimating $\mathbf{V}$, all embedding vectors are clustered into $S$ clusters using K-means. The masks are then constructed as follows
\begin{equation}
\hat{M}_{s,tf}=
\begin{cases}
      1, & \text{if}\ \mathbf{v}_{tf} \in c_s \\
      0, & \text{otherwise}
    \end{cases},
\end{equation}
with $c_s$ a cluster from K-means. \eqref{eq:spec_est} can then be used to estimate the STFT of the original source signals.




\subsection{i-vectors}
A speaker representation that is often used for speaker identification tasks is the i-vector \citep{dehak2011front,glembek2011simplification}.
To obtain such an i-vector, first a universal background model, based on a Gaussian mixture model (GMM-UBM), is trained on development data. A supervector $\mathbf{s}$ is derived for each utterance, using the UBM, by stacking the speaker-adapted Gaussian mean vectors. $\mathbf{s}$ is then represented by an i-vector $\mathbf{w}$ and its projection based on the total variability space,
\begin{equation}
\mathbf{s} \approx \mathbf{m}+\mathbf{T}\mathbf{w},
\end{equation}
where $\mathbf{m}$ is the UBM mean supervector, $\mathbf{w}$ is the total variability factor or i-vector and $\mathbf{T}$ is a low-rank matrix spanning a subspace with important variability in the mean supervector space and is trained on development data \citep{dehak2011front,glembek2011simplification}.

 If such i-vectors are explicitly presented to the input of the DC network, possibly less information would have to be retained in the LSTM memory as there is no need for an intrinsic speaker characterization \citep{zegers2017improving,zegers2018memory}. It is noteworthy that \cite{drude2018deep} managed to show separation quality improvement by adding an auxiliary speaker identification loss to the separation loss.
 
\section{Memory reset LSTM}
\label{sec:reset}
In this section we give a short summary of the LSTM before we describe the memory reset LSTM cell. Next we describe how this memory reset LSTM cell can be used in an RNN. Derivations are given in \ref{app:bewijs}. Finally, we discuss how computational costs can be reduced by using the grouped memory reset approach.

\subsection{Regular LSTM}

\begin{figure}
\centering
\includegraphics[width=0.95\linewidth]{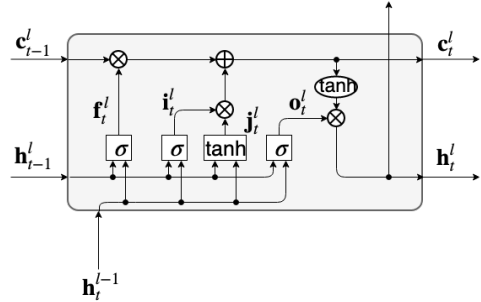}
\caption{Schematic of an LSTM cell. Based on \cite{blog}.}
\label{fig:LSTM}
\end{figure}

\begin{figure}
\centering
\includegraphics[width=0.85\linewidth]{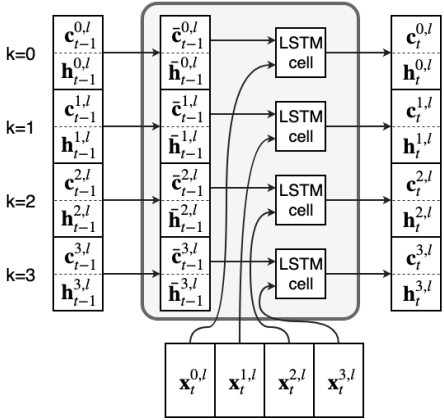}
\caption{Schematic of a memory reset LSTM cell with $T_{\text{reset}}=4$. It keeps $K=4$ instances of the hidden unit and cell state, which can be reset according to \eqref{eq:reset}. The LSTM cell is used to process each instance.}
\label{fig:reset_LSTM}
\end{figure}
  

The regular LSTM cell is shown in figure \ref{fig:LSTM} and is defined in \eqref{eq:LSTM_f}--\eqref{eq:LSTM_h}.
\begin{equation}
\label{eq:LSTM_f}
\mathbf{f}_t^l=\sigma(\mathbf{W}_f^l \mathbf{x}_t^{l} + \mathbf{R}_f^l \mathbf{h}_{t-1}^l + \mathbf{b}_f),
\end{equation}
\begin{equation}
\label{eq:LSTM_i}
\mathbf{i}_t^l=\sigma(\mathbf{W}_i^l \mathbf{x}_t^{l} + \mathbf{R}_i^l \mathbf{h}_{t-1}^l + \mathbf{b}_i),
\end{equation}
\begin{equation}
\label{eq:LSTM_o}
\mathbf{o}_t^l=\sigma(\mathbf{W}_o^l \mathbf{x}_t^{l} + \mathbf{R}_o^l \mathbf{h}_{t-1}^l + \mathbf{b}_o),
\end{equation}
\begin{equation}
\label{eq:LSTM_j}
\mathbf{j}_t^l=\tanh(\mathbf{W}_j^l \mathbf{x}_t^{l} + \mathbf{R}_j^l \mathbf{h}_{t-1}^l + \mathbf{b}_j),
\end{equation}
\begin{equation}
\label{eq:LSTM_c}
\mathbf{c}_t^l=\mathbf{c}_{t-1}^l \odot \mathbf{f}_t^l + \mathbf{j}_t^l \odot \mathbf{i}_t^l,
\end{equation}
\begin{equation}
\label{eq:LSTM_h}
\mathbf{h}_t^l= \tanh(\mathbf{c}_{t}^l) \odot \mathbf{o}_t^l,
\end{equation}
with $\mathbf{x}_t^l$, $\mathbf{c}_t^l$ and $\mathbf{h}_t^l$ the cell's input, state and output (or hidden unit) respectively, at time $t$ for layer $l=1,\ldots,L$
with $L$ the number of layers in the network. $\mathbf{f}_t^l$, $\mathbf{i}_t^l$ and $\mathbf{o}_t^l$ are called the forget gate, the input gate and the output gate, respectively. The input of an LSTM cell is the output from the layer below.
\begin{equation}
\label{eq:LSTM_in}
\mathbf{x}_t^l= \mathbf{h}_t^{l-1},
\end{equation}
for $l=2,\ldots,L$. The first layer receives the input from the LSTM-RNN. The output of the LSTM-RNN is the output from the last layer
\begin{equation}
\label{eq:LSTM_out}
\mathbf{h}_t= \mathbf{h}_t^{L}.
\end{equation}
An LSTM-RNN can be made bidirectional. In that case the backward direction processes the input from end to start. The $t-1$ subscript in \eqref{eq:LSTM_f}--\eqref{eq:LSTM_c} is replaced by $t+1$. We use $\overrightarrow{\mathbf{\bullet}}$ to denote the hidden units in the forward direction and $\overleftarrow{\mathbf{\bullet}}$ for the backward direction. There are two ways to combine the outputs from the forward and backward direction: either after every layer or only after the last layer. If the latter is chosen \eqref{eq:LSTM_in} becomes
\begin{equation}
\label{eq:BLSTM_in_sep}
\overrightarrow{\mathbf{x}}_t^l= \overrightarrow{\mathbf{h}}_t^{l-1} \qquad \qquad \overleftarrow{\mathbf{x}}_t^l= \overleftarrow{\mathbf{h}}_t^{l-1}.
\end{equation}
If inputs are instead combined after every layer we get
\begin{equation}
\label{eq:BLSTM_in_nonsep}
\overrightarrow{\mathbf{x}}_t^l=\left(
                                    \begin{array}{c}
                                        \overrightarrow{\mathbf{h}}_{t}^{l-1} \\ 
                                        \overleftarrow{\mathbf{h}}_{t}^{l-1}  
                                    \end{array}
                                \right)
\qquad
\overleftarrow{\mathbf{x}}_t^l=\left(
                                    \begin{array}{c}
                                        \overrightarrow{\mathbf{h}}_{t}^{l-1} \\ 
                                        \overleftarrow{\mathbf{h}}_{t}^{l-1}  
                                    \end{array}
                                \right).
\end{equation}
In both cases, for the final output of the network, \eqref{eq:LSTM_out} becomes
\begin{equation}
\label{eq:BLSTM_out}
\mathbf{h}_t=\left(
                                    \begin{array}{c}
                                        \overrightarrow{\mathbf{h}}_{t}^{L} \\ 
                                        \overleftarrow{\mathbf{h}}_{t}^{L}  
                                    \end{array}
                                \right).
\end{equation}

In our experiments we found that it was better to combine outputs after every layer, even if the total number of trainable parameters was kept unchanged.

\subsection{Memory reset LSTM cell}

To limit the recurrent information, the cell state $\mathbf{c}_t^l$ and hidden unit $\mathbf{h}_t^l$ will be reset using a fixed reset period $T_{\text{reset}}$. This assures that only information of the last $T_{\text{reset}}$ frames can be used (or $T_{\text{reset}}-1$ context frames). However, the time frame after such a reset would see no recurrent information at all. In fact, at every time step, an output should be produced based on $T_{\text{reset}}$ frames of information. To achieve this, $K=T_{\text{reset}}$ different instances of the cell state and hidden unit are kept, each reset at different moments in time. The instance that will be reset at time $t$ is the instance $k^*_t$ for which

\begin{equation}
\label{eq:kstar}
    k^*_t=t\ \text{mod}\ K,
\end{equation}
with $t = 0, \ldots, T-1$.

The reset operation can be formulated as
\begin{equation}
\label{eq:reset}
\left(\bar{\mathbf{h}}_{t-1}^{k,l},\bar{\mathbf{c}}_{t-1}^{k,l}\right)=
\begin{cases}
      \left(\mathbf{0},\mathbf{0}\right), & \text{if}\ k=k^*_t\ \\
      \left(\mathbf{h}_{t-1}^{k,l},\mathbf{c}_{t-1}^{k,l}\right), & \text{otherwise}
    \end{cases},
\end{equation}
with $k = 0, \ldots, K-1$.
\eqref{eq:LSTM_f}--\eqref{eq:LSTM_h} are updated to \eqref{eq:reset_LSTM_simple_f}--\eqref{eq:reset_LSTM_simple_h}.
\begin{equation}
\label{eq:reset_LSTM_simple_f}
\mathbf{f}_t^{k,l}=\sigma(\mathbf{W}_f^l \mathbf{x}_t^{k,l} + \mathbf{R}_f^l \bar{\mathbf{h}}_{t-1}^{k,l} + \mathbf{b}_f),
\end{equation}
\begin{equation}
\label{eq:reset_LSTM_simple_i}
\mathbf{i}_t^{k,l}=\sigma(\mathbf{W}_i^l \mathbf{x}_t^{k,l} + \mathbf{R}_i^l \bar{\mathbf{h}}_{t-1}^{k,l} + \mathbf{b}_i),
\end{equation}
\begin{equation}
\label{eq:reset_LSTM_simple_o}
\mathbf{o}_t^{k,l}=\sigma(\mathbf{W}_o^l \mathbf{x}_t^{k,l} + \mathbf{R}_o^l \bar{\mathbf{h}}_{t-1}^{k,l} + \mathbf{b}_o),
\end{equation}
\begin{equation}
\label{eq:reset_LSTM_simple_j}
\mathbf{j}_t^{k,l}=\tanh(\mathbf{W}_j^l \mathbf{x}_t^{k,l} + \mathbf{R}_j^l \bar{\mathbf{h}}_{t-1}^{k,l} + \mathbf{b}_j),
\end{equation}
\begin{equation}
\label{eq:reset_LSTM_simple_c}
\mathbf{c}_t^{k,l}=\bar{\mathbf{c}}_{t-1}^{k,l} \odot \mathbf{f}_t^{k,l} + \mathbf{j}_t^{k,l} \odot \mathbf{i}_t^{k,l},
\end{equation}
\begin{equation}
\label{eq:reset_LSTM_simple_h}
\mathbf{h}_t^{k,l}= \tanh(\mathbf{c}_{t}^{k,l}) \odot \mathbf{o}_t^{k,l}.
\end{equation}
A visualization for $K=4$ is given in figure \ref{fig:reset_LSTM}.

\subsection{Deep memory reset LSTM-RNN}
\label{sec:reset_rnn}
\begin{figure*}
  \centering
  \includegraphics[width=\linewidth]{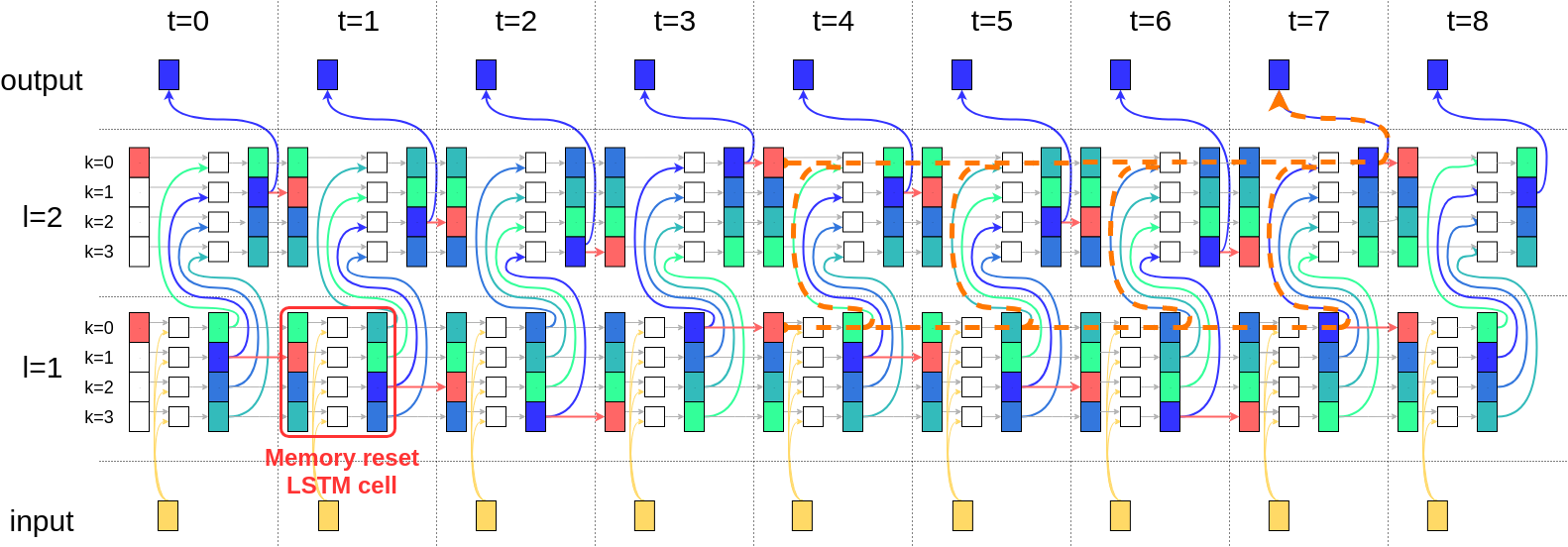}
  \caption{Example of a unidirectional memory reset LSTM-RNN with $L=2$, $K=4$ and $T=9$. Color coding is according to \eqref{eq:tau_def}. An instance is colored red when it is reset ($k=k^*_t$, according to \eqref{eq:kstar}). Connections between layers are according to \eqref{eq:reset_LSTM_in}. The final output follows \eqref{eq:reset_LSTM_out}. When following the orange dashed line, indicating the data dependencies, backward, one can verify that indeed exactly $K$ frames are used to produce an output.}
  \label{fig:L2_reset}
\end{figure*}

For multi-layer memory reset LSTM-RNNs, instances need input from instances of the layer below. In other words, an equivalent for \eqref{eq:LSTM_in} has to be found for the memory reset LSTM-RNN. We introduce a new variable $\tau_t^{k,l}$ which is equal to how long ago instance $k$ of layer $l$ was last reset (or how many context frames instance $k$ of layer $l$ considered at time $t$). \eqref{eq:tau_def_proof} shows that\footnote{The context is restricted at the edges of the data sequence. $\tau_t^{k,l}$ can never exceed $t$. This is not a restriction of the memory reset approach but intrinsic to the data sequence.}
\begin{equation}
\label{eq:tau_def}
    \tau_t^{k,l} = (t-k)\ \text{mod}\ K.
\end{equation}
The value of $\tau_t^{k,l}$ is color coded in figure \ref{fig:L2_reset} for $K=4$. If an instance is colored light green, then $\tau_t^{k,l}=0$. If an instance is colored dark blue, then $\tau_t^{k,l}=3\ (=K-1)$. An instance should receive input from the instance of the layer below with the same number of context frames $\tau_t^{k,l}$. The orange dashed line in figure \ref{fig:L2_reset} shows that this way no information further than $K$ frames can be used. \eqref{eq:uni_connection_proof} shows that for a unidirectional memory reset LSTM-RNN, this is obtained when instance $k$ receives input from instance $k$ from the layer below. \eqref{eq:LSTM_in} generalizes to
\begin{equation}
    \label{eq:reset_LSTM_in}
    \mathbf{x}_t^{k,l} = \mathbf{h}_t^{k,l-1}.
\end{equation}
We introduce a new simplified notation $k' \leftarrow k''$, stating that instance $k'$ of layer $l$ receives input from instance $k''$ of layer $l-1$. In this notation \eqref{eq:reset_LSTM_in} becomes $k \leftarrow k$.
The final output of the network at time $t$ is the instance with the maximum number of context frames at that time. This is the instance that will be reset at time $t+1$ (see \eqref{eq:reset_LSTM_out_proof}). \eqref{eq:LSTM_out} generalizes to
\begin{equation}
\label{eq:reset_LSTM_out}
\mathbf{h}_t= \mathbf{h}_t^{k^*_{t+1},L}.
\end{equation}

\begin{figure*}
  \centering
  \includegraphics[width=\linewidth]{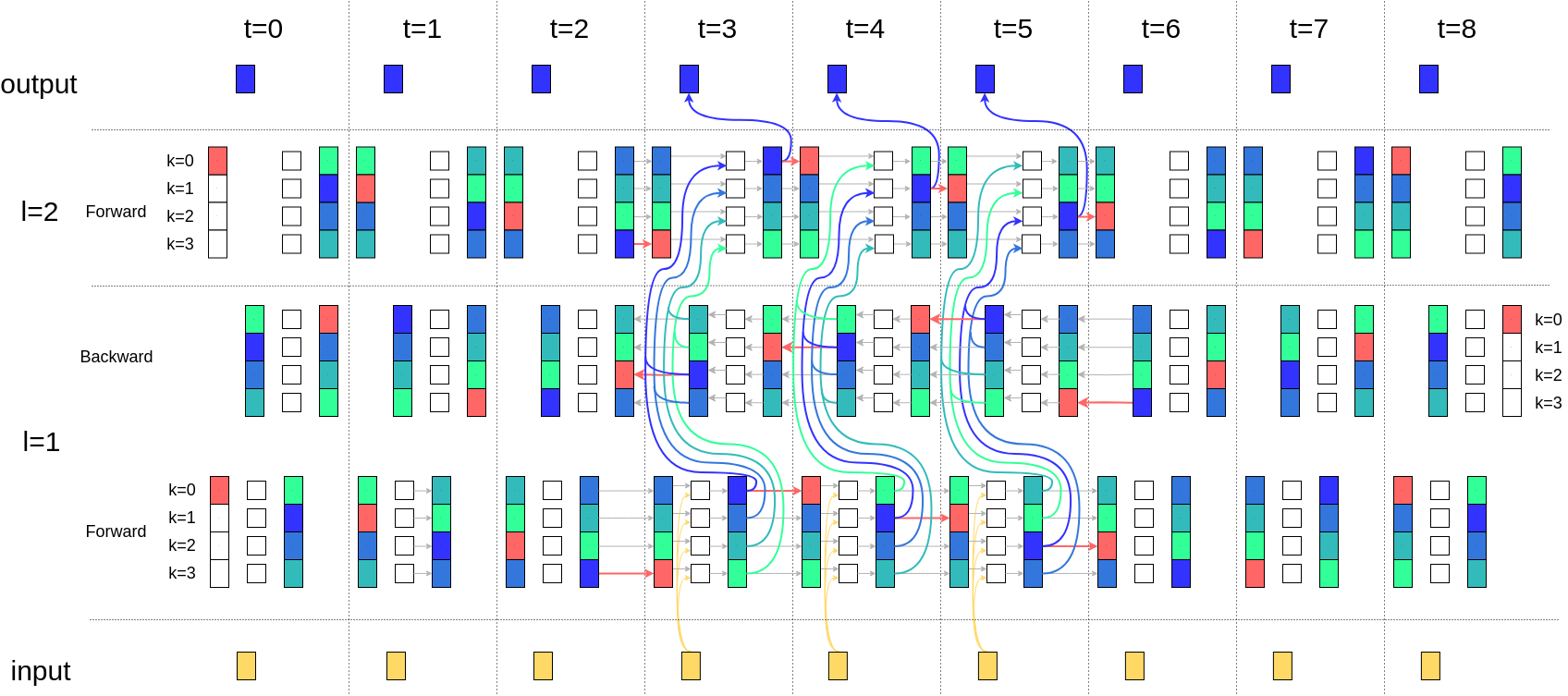}
  \caption{Similar to figure \ref{fig:L2_reset} but for bidirectional memory reset LSTM-RNN. To prevent cluttering of the image, only the forward direction of the second layer is shown and connections are only drawn for $t=\{3,4,5\}$.}
  \label{fig:bidir_L2_reset}
\end{figure*}
For bidirectional LSTM-RNNs we apply the same equal context method to find the following connections, replacing \eqref{eq:BLSTM_in_nonsep}
\begin{equation}
    \label{eq:bi_connection_for}
    \overrightarrow{k} \leftarrow \left(
                                    \begin{array}{c}
                                        \overrightarrow{k} \\ 
                                        ((T-1)-2t+\overleftarrow{k})\ \text{mod}\ K
                                    \end{array}
                                \right),
\end{equation}
\begin{equation}
    \label{eq:bi_connection_back}
    \overleftarrow{k} \leftarrow \left(
                                    \begin{array}{c}
                                        (-(T-1)+2t+\overrightarrow{k})\ \text{mod}\ K \\ 
                                        \overleftarrow{k}  
                                    \end{array}
                                \right).
\end{equation}
\eqref{eq:bi_connection_for}-\eqref{eq:bi_connection_back} are derived in \eqref{eq:bi_connection_for_proof}-\eqref{eq:bi_connection_back_proof} and can be verified in figure \ref{fig:bidir_L2_reset}.

The reset period $T_{\text{reset}}$ needs not be the same for every layer. For instance, we could allow the lower levels to operate on short-term information and let the higher layers cope with the long-term dependencies. By connecting an instance with the correct instance of the previous layer at every time $t$, we can still make sure the the number of context frames per layer is limited to a chosen value. \eqref{eq:bi_connection_for}-\eqref{eq:bi_connection_back} generalize to \eqref{eq:bi_connection_for_diffK}-\eqref{eq:bi_connection_back_diffK}.

\subsection{Grouped memory reset LSTM}
\label{sec:groups}
\begin{figure*}
  \centering
  \includegraphics[width=\linewidth]{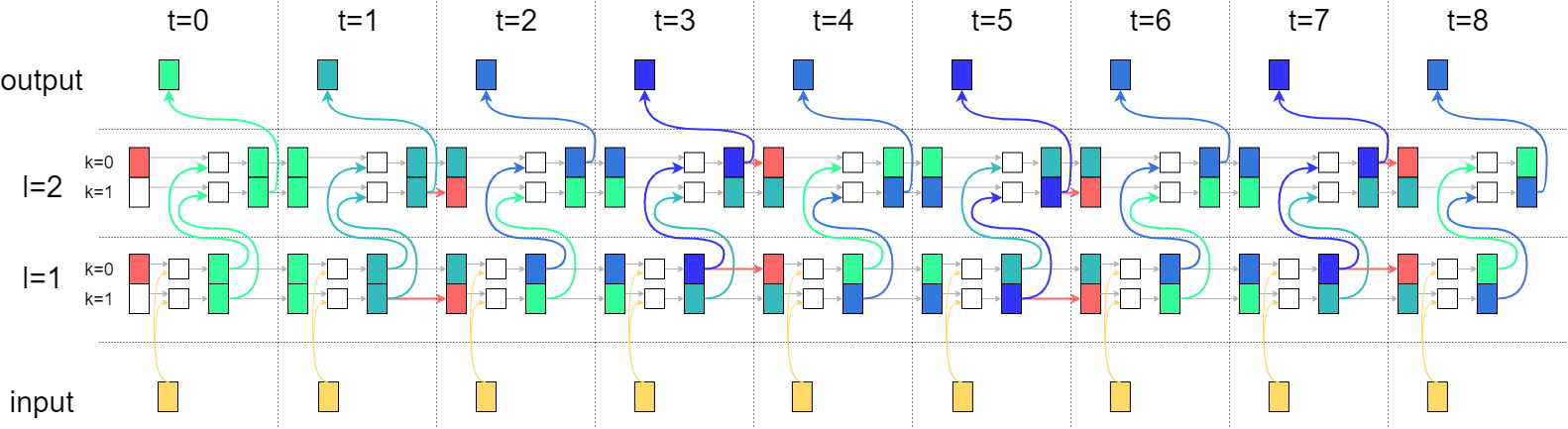}
  \caption{Example of a unidirectional memory reset LSTM-RNN using groups with $L=2$, $K=2$, $G=2$ and $T=9$. $T_{\text{reset}}=4$, just as in figure \ref{fig:L2_reset}. Color coding is according to \eqref{eq:tau_def_proof_groups}. An instance is colored red when it is reset ($k=k^*_t$, according to \eqref{eq:kstar_group}).}
  \label{fig:L2_groups_reset}
\end{figure*}
Until now, a different instance is reset at every time step such that each instance is reset every $K$ time steps. The number of instances $K$ is therefore equal to the reset period $T_{\text{reset}}$ and the computational requirements grow as the reset period (or the memory span) becomes larger.
By using the reset operation only every $G$ time steps, an instance will only be reset every $KG$ time steps and thus the reset period becomes $T_{\text{reset}}=KG$. Then the number of instances can be reduced with a factor $G$, for the same $T_{\text{reset}}$. 
\eqref{eq:kstar} is changed to

\begin{equation}
\label{eq:kstar_group}
\begin{cases}
    k^*_t=(t/G)\ \text{mod} \ K ,&\text{if} \ t \equiv 0\ (\text{mod}\ G)\\
    \text{no reset} & \text{otherwise} \\
\end{cases}
\end{equation}

A visualization of the grouped memory reset approach is given in figure \ref{fig:L2_groups_reset}. The downside is that instead of allowing the LSTM to use $T_{\text{reset}}=KG$ frames of input, it will use between $KG-(G-1)$ and $KG$ frames of input, as shown in \eqref{eq:gamma_group_max_L_for}-\eqref{eq:gamma_group_max_L_back} (also see output layer in figure \ref{fig:L2_groups_reset}). However this need not be a concern, since the computational problems arise only for large number of instances and then $KG>>G$. A similar approach was taken by \cite{el1996hierarchical} where \textit{frame grouping} was used to reduce computational burden. Derivation for the connections between grouped memory reset LSTM layers is given in \ref{app:bewijs_grouped}.

\section{Experimental setup}
\label{sec:exp}
For the MSSS task, mixtures of two speakers were used from the corpus introduced by \cite{hershey2016deep}. These mixtures were artificially created by mixing single speaker utterances from the Wall Street Journal 0 (WSJ0) corpus. A gain for the first speaker compared to the second speaker was randomly chosen between 0 and 5 dB. Utterances were sampled at 8kHz and the length of the mixture was chosen equal to the shortest utterance in the mixture to maximize the overlap. The training and validation sets contained 20,000 and 5,000 mixtures, respectively from 101 speakers, while the test set contained 3,000 mixtures from 16 held-out speakers. A STFT with a 32 ms window length and a hop size of 8 ms were used, so the context span is defined as $T_{\text{span}}=(T_{\text{reset}}-1)*8\text{ms}$. 
Notice that for bidirectional networks, this memory span is used for both the left and right context. Performance for MSSS is measured by the average signal-to-distortion ratio (SDR) improvements on the test set, using the \texttt{bss\_eval} toolbox \citep{vincent2006performance}. In the experiments we make a distinction between male-female mixtures and same gender mixtures. The former is regarded much easier than the latter. 

The memory reset approach was applied to a network of two fully connected bidirectional LSTM-RNN layers, with 600 hidden units each. The reset is applied in both directions of the network, unless stated otherwise. Hidden units of both directions are concatenated before being passed to the next layer, as was expressed by \eqref{eq:BLSTM_in_nonsep}. For DC the embedding dimension was chosen at $D=20$ and since the frequency dimension was $F=129$, the total number of output nodes was $DF=20*129=2580$. Curriculum learning was applied by first training the networks on 100 frame segments, before training on the full mixture \citep{bengio2009curriculum,hershey2016deep}. The weights and biases were optimized with the Adam learning algorithm \citep{kingma2014adam} and early stopping on the validation set was used. The log-magnitude of the STFT coefficients were used as input features and were mean and variance normalized. Zero mean Gaussian noise with standard deviation $0.2$ was applied to the training features to avoid local optima. When estimating the performance of a network for a certain reset frequency $T_{\text{reset}}$, always two networks, with different initializations, were trained an tested to cope with variance on the evaluated performance. When $T_{\text{reset}}=\infty$, a regular LSTM-RNN instead of a memory reset LSTM-RNN was used. All networks were trained using TensorFlow \citep{abadi2016tensorflow} and the code for all the experiments can be found online\footnote{\texttt{github.com/JeroenZegers/Nabu-MSSS}}.

For the i-vectors, a UBM and $T$ matrix were trained on the Wall Street Journal 1 (WSJ1) corpus, using the \texttt{MATLAB MSR Identity Toolbox v1.0} \citep{sadjadi2013msr}. 13-dimensional mel-frequency cepstral coefficients (MFCCs) were used as features, the UBM had 256 Gaussian mixtures and the i-vectors were 10-dimensional, as was done by \cite{zegers2017improving}. The i-vectors used in the experiments were obtained from the original single speaker utterances of WSJ0 but could also be obtained from speech signal reconstructions after source separation, as was done by \cite{zegers2017improving}. The former was chosen since it provides a cleaner speaker representation.

\section{Discussion}
\label{sec:res}
We use the memory reset LSTM-RNN to gain insights in the importance of the memory span of the LSTM on the task performance. The first experiment (section \ref{sec:res_groups}) is solely to verify that indeed for large $T_{\text{reset}}$ we can allow $G>1$ and still measure the memory effect correctly. This allows us to use the grouping technique in the other experiments for computational efficiency. The next experiment (section \ref{sec:res_ivecs}) analyses the importance of different memory spans, with and without speaker information. This gives insights into what the LSTM tries to remember. The third experiment (section \ref{sec:layer_reset}) uses a short memory span for the first layer and a large one for the second layer. We find little difference with a network where both layers have large memory spans. This confirms the existence of hierarchy in memorization. Finally, in section \ref{sec:for_back_reset} we look at the differences between the effect of the forward memory and the backward memory on the task performance.

\subsection{Verification of grouping technique}
\label{sec:res_groups}

\begin{figure*}
  \centering
  \includegraphics[width=\linewidth]{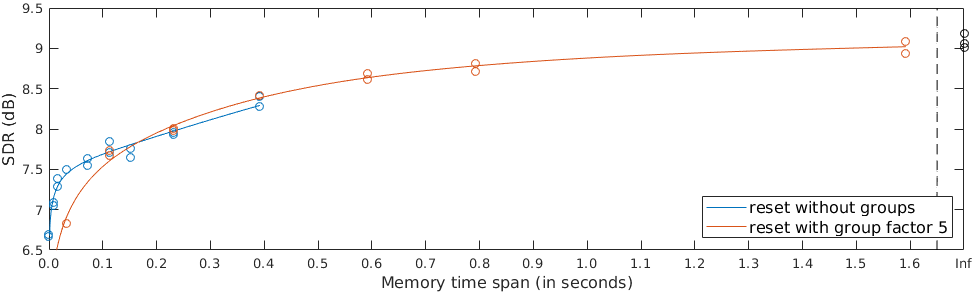}
  \caption{Average separation results for networks trained and evaluated for different memory time spans. The blue curve uses no grouping ($G=1$), the orange curve uses a grouping factor of 5 ($G=5$). Every experiment was performed twice to cope with variance on the evaluated performance.}
  \label{fig:proof_groups}
\end{figure*}

In figure \ref{fig:proof_groups} the separation performance for networks with different memory time spans (without grouping) are given in blue. Notice that no networks were trained for $T_{\text{span}}>400\text{ms}$ since the computational memory requirements became too large for $K>50$. In orange, a group factor of $G=5$ ($=40 \text{ms}$) was applied. 

It is clear that a group factor of $G=5$ can be used for $T_{\text{span}}>100\text{ms}$ (since $100\text{ms} >> 40\text{ms}$), without loss of performance. Thus, the grouping method is a valid way to break the linear dependence of the computational memory requirements on the reset period for large reset periods. In the remainder of the paper, results for $T_{\text{span}}\leq400\text{ms}$ will be given without the grouping approach and results for $T_{\text{span}}>400\text{ms}$ will be given with a reset factor of $G=5$.

\subsection{Memory span with and without speaker information}
\label{sec:res_ivecs}
\begin{figure*}
  \centering
  \includegraphics[width=\linewidth]{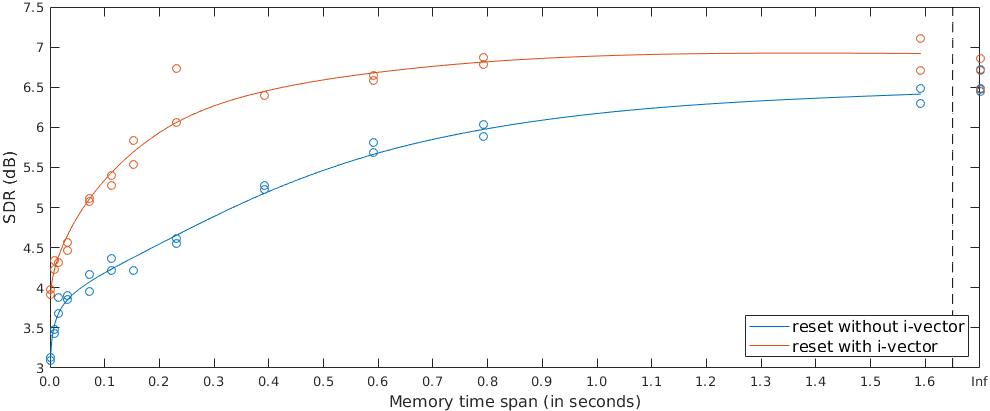}
  \caption{Average separation results for networks using different memory time spans, evaluated on same gender mixtures. In blue, the results are given without using i-vectors. In orange, the results when the i-vector of both speakers are appended to the input of the network.}
  \label{fig:same_gen}
\end{figure*}

\begin{figure*}
  \centering
  \includegraphics[width=\linewidth]{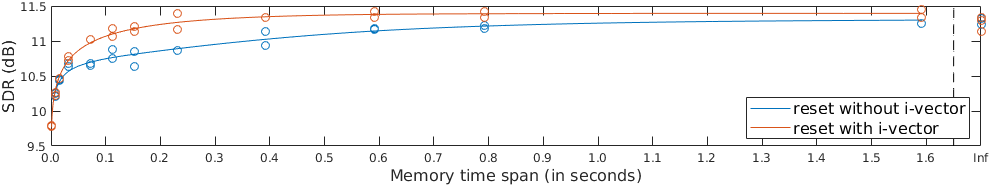}
  \caption{Average separation results for networks using different memory time spans, evaluated on male-female mixtures. In blue, the results are given without using i-vectors. In orange, the results when the i-vector of both speakers are appended to the input of the network.}
  \label{fig:diff_gen}
\end{figure*}

Figure \ref{fig:same_gen} shows the average separation performance for same gender (male-male and female-female) mixtures, with and without the i-vectors of both speakers appended to the input. Figure \ref{fig:diff_gen} shows the male-female results for the same networks. Since the results are clearly different, the figures will be discussed separately.

In figure \ref{fig:same_gen} the blue curve quickly rises when the memory span is extended from 0ms to 400ms. This effect is also noted for the models where i-vectors were appended to the input (orange curve). Here, the increase in performance cannot be explained by a better speaker characterization, since the information is already present in the i-vectors. Therefore, the increase in performance of the blue curve cannot solely be explained by a better speaker characterization.
The separation task seems to take phonetic information (about 100ms \citep{gay1968effect,umeda1975vowel}) into account. Features like common onset, common offset and harmonicity playing a central role in auditory grouping in humans \citep{bregman1994auditory} are compatible with this observed time scale as well. Information spanning several 100ms also seems important. In this range, effects like phonotactics, lexicon and prosody can play a role but further research is necessary to determine to which extent each of these are individually important for MSSS. In \cite{appeltans2019practical} it was found that models trained on one language generalize to some extent to a different language, making it unlikely that lexical information is key for MSSS.
If the memory span is restricted to $T_{\text{span}}<400\text{ms}$, there is a clear difference between the networks with and without i-vectors at the input. If the memory is restricted, it is difficult to characterize and separate speakers with the same gender. Using i-vectors helps to solve this problem. 

For $T_{\text{span}}>400\text{ms}$, the SDR without i-vectors keeps increasing with increasing time span, while the curve with i-vectors remains approximately flat. This indicates that when further increasing the memory span at this point, the LSTM-RNN without i-vector can only learn to model the speakers better. 
On the other hand, grammatical information, which is expected to span much longer than $400\text{ms}$ \citep{miller1984articulation}, is not considered in the LSTM for the MSSS task.
\cite{mohamed2015deep} found that performance for their ASR task did not improve for $T_{\text{span}}>250\text{ms}$ (500 ms is mentioned, but this includes both left and right context). We notice that this is not the case for MSSS and conclude that the need for a longer time span is mainly caused by the subtask of speaker characterization.



Both curves for the male-female mixtures (figure \ref{fig:diff_gen}) quickly converge to the result without memory restrictions. There is also a limited difference between results with and without i-vectors at the input, indicating that it is indeed easy to distinguish a male from a female speaker. The result for $T_{\text{span}}=0\text{ms}$ is far above the optimal result for same gender mixtures (figure \ref{fig:same_gen}). Instantaneous pitch and formant information seems to achieve most of the effect. Male and female speakers are easily separable, even without any context. Since we are interested in how the LSTM-RNN uses this context, we will only report same gender separation results in the remainder of this paper. However, we do observe that while there is no significant difference between the reset with and without i-vectors for $T_{\text{span}}<30\text{ms}$, there is a slight improvement to be found when including the i-vectors for $T_{\text{span}}>30\text{ms}$.  This might indicate that most different-gender mixtures can be separated based on local information (pitch, formants), while for some cases, more sophisticated speaker characterization at longer time span is required. In the latter case, unsurprisingly, i-vectors help.


\subsection{Layer wise reset}
\label{sec:layer_reset}

\begin{figure}
\centering
\includegraphics[width=1.0\linewidth]{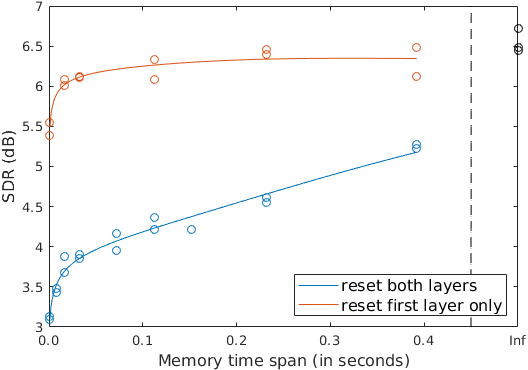}
\caption{Average separation results for networks using different memory time spans, evaluated on same gender mixtures. The blue curve is identical to the blue curve in figure \ref{fig:same_gen}. The orange curve shows the result when memory reset is only applied to the first layer and the second layer has no memory restrictions. The results at infinity, colored in black, use no memory reset.
}
\label{fig:first_layer_only}

\end{figure}

The orange curve in figure \ref{fig:first_layer_only} shows the performance when memory reset is only applied to the first layer of the network. Naturally, performance is better compared to resetting both layers. However, it is interesting to note that optimal performance is already approximately achieved with a memory time span of less than 50ms. This confirms the hypothesis by \cite{chung2016hierarchical} that it is sufficient to allow larger time spans only in the deeper layers to model the higher level abstractions.

\subsection{Forward and backward reset}
\label{sec:for_back_reset}
\begin{figure}
\centering
\includegraphics[width=1.0\linewidth]{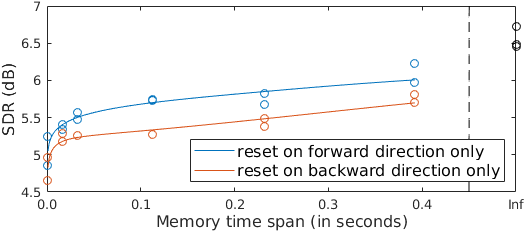}
\caption{Average separation results for networks using different memory time spans, evaluated on same gender mixtures. The blue curve shows the result when memory reset is only applied to the forward direction and the backward direction layer has no memory restrictions. The orange curve applies memory reset only to the backward direction. The results at infinity, colored in black, use no memory reset.}
\label{fig:reset_per_dir}
\end{figure}

To our knowledge, there has been very little analysis on the relative importance between the forward direction and backward direction of a bidirectional RNN. Early results for a forward-only and backward-only model are given in \cite{schuster1997bidirectional} and \cite{graves2005framewise}.
Figure \ref{fig:reset_per_dir} shows the difference in performance between a bidirectional LSTM-RNN when memory reset is applied only on the forward direction (blue curve) compared to only on the backward direction (orange curve). For the blue curve, the networks evaluated at $T_{\text{span}}=0\text{ms}$, essentially correspond to a backward-only RNN. As $T_{\text{span}}$ is increased, more forward information is allowed but the backward direction remains dominant since it has no memory restrictions. It is noted that at $T_{\text{span}}=0\text{ms}$, the backward-only LSTM-RNN slightly outperforms the forward-only LSTM-RNN. This small but consistent difference is kept as the time span for the non-dominant direction increases to 400ms. 

\begin{figure}
\centering
\includegraphics[width=0.95\linewidth]{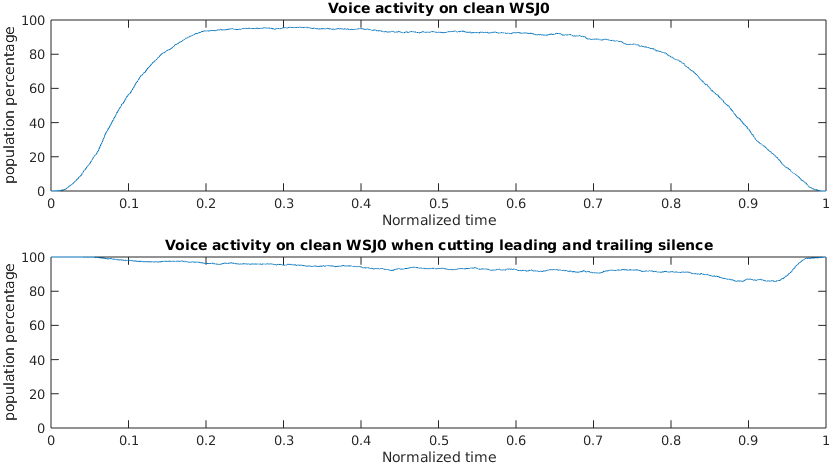}
\caption{Speech activity percentage of clean WSJ0 utterance when audio length normalized is to 1. For the original utterances (top) and when cutting leading and trailing silence (bottom).}
\label{fig:wsj_vad}
\end{figure}

\begin{figure}
\centering
\includegraphics[width=0.95\linewidth]{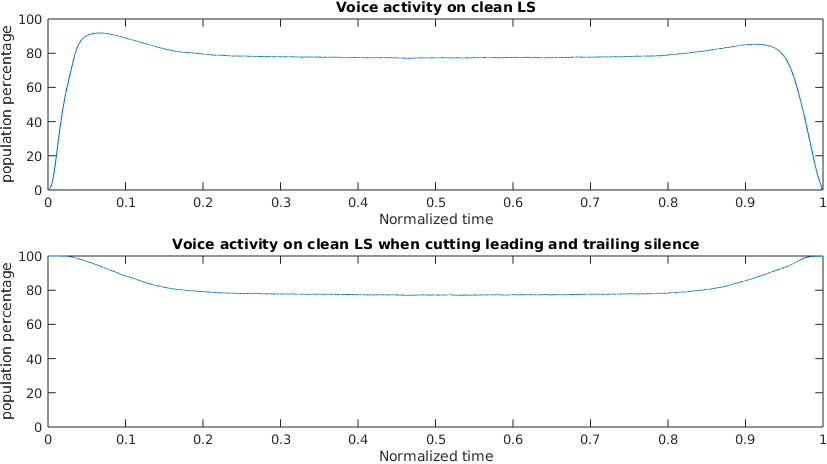}
\caption{Speech activity percentage of clean LS utterance when audio length is normalized to 1. For the original utterances (top) and when cutting leading and trailing silence (bottom).}
\label{fig:ls_vad}
\end{figure}

\begin{figure}
\centering
\includegraphics[width=1.0\linewidth]{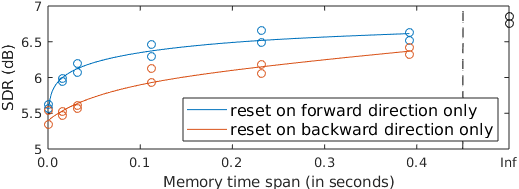}
\caption{Similar to Figure \ref{fig:reset_per_dir}, but on the LS mixture dataset.}
\label{fig:ls_reset_per_dir}
\end{figure}

While looking for reasons that could explain this difference, we found that speakers in WSJ0 ended their utterance with \enquote{period}, often taking a short break before pronouncing it\footnote{For instance, in the \nth{4} CHiME challenge \citep{vincent20164th} this part is removed from the utterance.}. This leads to some asymmetry in the speech activity as is shown in Figure \ref{fig:wsj_vad}, when measuring with a voice activity detector (VAD) \citep{tan2010low}. After combining single speaker utterances to mixtures, this leads to less overlapping speech near the end of the mixture and might explain the difference we observe in Figure \ref{fig:reset_per_dir}. Furthermore, as \enquote{period} is pronounced at the end of every utterance, it might behave as a prompt for text-dependent speaker recognition \citep{variani2014deep}. 
To exclude these unwanted effects, the \emph{LibriSpeech} (LS) dataset \citep{panayotov2015librispeech}, which does not contain verbal punctuation, was used to artificially create mixtures\footnote{This dataset has also been used by other papers for MSSS \citep{stephenson2017monaural,mobin2018convolutional}}. To ensure symmetry in speech activity, leading and trailing silence in the single speaker utterances were cut (see Figure \ref{fig:ls_vad} bottom). The forward-backward experiment was repeated on the newly created dataset and results are shown in Figure \ref{fig:ls_reset_per_dir}. We see a similar trend as for Figure \ref{fig:reset_per_dir} and retain our conclusion that the backward direction is slightly more important than the forward direction for MSSS.


However, the question of what causes this difference remains unanswered. 
It seems to suggest that cues in speech for MSSS are partly asymmetric.
It has been found that voice onset time (VOT) is a predictive cue for post-aspiration \citep{klatt1975voice,LiskerLeigh1967SEoC}, while similar conclusions have been drawn for voice offset time (VoffT) \citep{SinghRita2016Trov,pind1996rate}. 
Furthermore, it has also been observed that there is an acoustical asymmetry in vowel production \citep{patel2017relationship}.
Finally, reverberation could also play a role in a realistic cocktail party scenario, but this is expected not to be relevant in our experiments, considering the recording setup for WSJ0 and LS.
We leave it to further research to indicate to which extend these asymmetric cues help in MSSS.

Comparing the result without memory restrictions (black colored results at $T_{\text{span}}=\infty$) with the backward constrained results (orange curve) gives an indication on the SDR drop when only limited backward data is available. This can be relevant for near real-time applications with limited allowed delay. We notice that online implementations with limited delay (shorter than 100ms) loose roughly 1.5dB in SDR for same gender mixtures, compared to offline implementations.





%

\section{Conclusions}
\label{sec:conc}
A memory reset approach was developed and applied to an LSTM-RNN to find the importance in different time spans for the task of MSSS. 
Short-term linguistic processes (time spans shorter than 100ms) have a strong impact on the separation performance. Above 400ms the network can only learn better speaker characterization and other effects like grammar are not considered by the LSTM

Furthermore, the reset method allowed us to verify that performance-wise it is sufficient to implement longer memory in deeper layers. Finally, we found that the backward direction is slightly more important than the forward direction for a bidirectional LSTM-RNN. 

The next step of this research would be to use the insights we have gained to adapt the architecture of the (LSTM-)RNN. We would like to encourage other researchers to apply a similar timing analysis for RNNs in their field. Either with the leaky approach (straightforward implementation, but no exact timings) or the memory reset or segment approach (less trivial implementation with higher computational burdens, but assuring exact timings). Moreover, these methods allow to assess the memory implications on the RNN for a certain subtask. For our task, the importance of speaker characterization was determined by comparing results with and without adding oracle i-vectors to the input of the network. This technique is generalizable to other tasks. For instance, in language modeling for French the gender of the subject must be remembered, possibly over many words, to conjugate the perfect tense accordingly. An oracle binary input (male/female) depending on the gender of the relevant subject could be provided. Comparing results with and without this additional binary input, could give an idea on the importance of this subtask on the memory of the RNN.

\appendix

\section{Inter-layer connections}
\label{app:bewijs}
\subsection{Unidirectional reset LSTM-RNN}
Given \eqref{eq:kstar}, instance $k$ of layer $l$ will be reset at times $t_{k,l}$ given by
\begin{equation}
\label{eq:t_for_k}
    t_{k,l} = k + \alpha K,
\end{equation}
with $\alpha$ a natural number. Therefore, the number of time steps $\tau_t^{k,l}$ between time $t$ and the last time instance $k$ was reset before time $t$ is given by
\begin{equation}
\begin{split}
    \label{eq:tau_def_proof}
    \tau_t^{k,l} &= (t-t_{k,l})\ \text{mod}\ K \\
    &= (t-k-\alpha K)\ \text{mod}\ K = (t-k)\ \text{mod}\ K.
\end{split}
\end{equation}
A time $t$ instance $k$ of layer $l$ contains $\tau_t^{k,l}$ frames of context. We would like this instance to receive input from an instance of the layer below with the same number of context frames $\tau_t^{k,l}$. In other words, we would like to find the instance that was reset at time $t-\tau_t^{k,l}$ in layer $l-1$. Using \eqref{eq:kstar} we find this to be
\begin{equation}
\begin{split}
\label{eq:uni_connection_proof}
    k^{*,l-1}_{t-\tau_t^{k,l}} &= (t-\tau_t^{k,l})\ \text{mod}\ K \\
    &= (t-(t-k)\ \text{mod}\ K)\ \text{mod}\ K \\
    &= k\ \text{mod}\ K = k.
\end{split}
\end{equation}
This simply means that instance $k$ from layer $l$ should receive input from instance $k$ from layer $l-1$. Or in simplified notation $k \leftarrow k$. Finally, the last layer of the LSTM-RNN should output a single instance which will be the final output of the network. We choose this to be the instance with the maximum number of context frames $\tau_t^{max,L} = K-1$. This means that the instance was reset at $t-\tau_t^{max,L} = t-K+1$. Thus the instance to select is
\begin{equation}
\begin{split}
\label{eq:reset_LSTM_out_proof}
    k^{*,L}_{t-K+1} &= (t-K+1)\ \text{mod}\ K \\
    &= (t+1)\ \text{mod}\ K = k^{*,L}_{t+1}.
\end{split}
\end{equation}

\subsection{Bidirectional reset LSTM-RNN}
For bidirectional LSTM-RNNs, we take a similar approach. 
Equivalent to \eqref{eq:kstar}, \eqref{eq:t_for_k} and \eqref{eq:tau_def_proof}, we define
\begin{equation}
    \overrightarrow{k}^{*,l}_t = t\ \text{mod}\ K_l,
\end{equation}
\begin{equation}
    t_{\overrightarrow{k},l} = \overrightarrow{k} + \alpha K_l,
\end{equation}
\begin{equation}
    \tau_t^{\overrightarrow{k},l} = (t-\overrightarrow{k})\ \text{mod}\ K_l.
\end{equation}
For the backward direction we find
\begin{equation}
    \overleftarrow{k}^{*,l}_t = ((T-1)-t)\ \text{mod}\ K_l,
\end{equation}
\begin{equation}
    t_{\overleftarrow{k},l} = (T-1) - (\overleftarrow{k} + \alpha K_l),
\end{equation}
\begin{equation}
\begin{split}
    \tau_t^{\overleftarrow{k},l} &= (t_{\overrightarrow{k},l}-t)\ \text{mod}\ K_l \\
    &= ((T-1) - t - \overleftarrow{k} - \alpha K_l)\ \text{mod}\ K_l \\
    &= ((T-1)-t-k)\ \text{mod}\ K_l.
\end{split}
\end{equation}
Again, we want an instance to receive input from an instance in the layer below with the same number of context frames. For the instances in the forward direction these are $\overrightarrow{k}^{*,l-1}_{t-\tau_t^{\overrightarrow{k},l}}$ and $\overleftarrow{k}^{*,l-1}_{t+\tau_t^{\overrightarrow{k},l}}$. When we use the same reset period for all layers ($K=K_l=K_{l-1}$), these are
\begin{equation}
\begin{split}
\label{eq:for_for_for_proof}
    \overrightarrow{k}^{*,l-1}_{t-\tau_t^{\overrightarrow{k},l}} &= (t-\tau_t^{\overrightarrow{k},l})\ \text{mod}\ K \\
    &= (t-(t-\overrightarrow{k})\ \text{mod}\ K)\ \text{mod}\ K \\
    &= \overrightarrow{k}\ \text{mod}\ K = \overrightarrow{k}.
\end{split}
\end{equation}
and
\begin{equation}
\begin{split}
\label{eq:back_for_for_proof}
    \overleftarrow{k}^{*,l-1}_{t+\tau_t^{\overrightarrow{k},l}}  &= ((T-1)-t-\tau_t^{\overrightarrow{k},l})\ \text{mod}\ K \\
     &= ((T-1)-t-(t-\overrightarrow{k})\ \text{mod}\ K)\ \text{mod}\ K \\
     &= ((T-1)-2t+\overrightarrow{k})\ \text{mod}\ K,
\end{split}
\end{equation}
respectively. As per \eqref{eq:BLSTM_in_nonsep}, in simplified notation this becomes
\begin{equation}
    \label{eq:bi_connection_for_proof}
    \overrightarrow{k} \leftarrow \left(
                                    \begin{array}{c}
                                        \overrightarrow{k} \\ 
                                        ((T-1)-2t+\overleftarrow{k})\ \text{mod}\ K
                                    \end{array}
                                \right).
\end{equation}
Similarly for the backward direction we find the inputs to be
\begin{equation}
    \label{eq:bi_connection_back_proof}
    \overleftarrow{k} \leftarrow \left(
                                    \begin{array}{c}
                                        (-(T-1)+2t+\overrightarrow{k})\ \text{mod}\ K \\ 
                                        \overleftarrow{k}  
                                    \end{array}
                                \right).
\end{equation}

As per \eqref{eq:BLSTM_out}, the final output of the network is a concatenation of the output of both directions of the last layer. We again choose these to be the instance with the maximum number of context frames $\tau_t^{\overrightarrow{max},L} = K-1$ and $\tau_t^{\overleftarrow{max},L} = K-1$ for the forward and backward direction, respectively. This means that corresponding instances were reset at $t-\tau_t^{\overrightarrow{max},L}=t-K+1$ and $t+\tau_t^{\overleftarrow{max},L}=t+K-1$. Thus the corresponding instances to select are

\begin{equation}
\begin{split}
\label{eq:reset_BLSTM_out_for_proof}
    \overrightarrow{k}^{*,L}_{t-K+1} &= (t-K+1)\ \text{mod}\ K \\
    &= (t+1)\ \text{mod}\ K = \overrightarrow{k}^{*,L}_{t+1}
\end{split}
\end{equation}
and
\begin{equation}
\begin{split}
\label{eq:reset_BLSTM_out_back_proof}
    \overleftarrow{k}^{*,L}_{t+K+1} &= ((T-1)-(t+K-1))\ \text{mod}\ K \\
    &= ((T-1)-(t-1))\ \text{mod}\ K \\
    &= \overleftarrow{k}^{*,L}_{t-1}.
\end{split}
\end{equation}
Thus \eqref{eq:BLSTM_out} generalizes to
\begin{equation}
\label{eq:reset_BLSTM_out_proof}
\mathbf{h}_t=\left(
                \begin{array}{c}
                    \overrightarrow{\mathbf{h}}_{t}^{\overrightarrow{k}^{*,L}_{t+1},L} \\ 
                    \overleftarrow{\mathbf{h}}_{t}^{\overleftarrow{k}^{*,L}_{t-1},L}  
                \end{array}
            \right).
\end{equation}

\subsection{Layer dependent reset period}
\label{app:layer_dep_reset}
If $K_l$ and $K_{l-1}$ are different, we would still like an instance to receive input from the layer below with the same number of context frames. However, this is not possible when the number of context frames exceeds the maximum number of context frames in the layer below (bounded by $K_{l-1}-1$). therefore, a new variable $\bar{\tau}$ is introduced, which is defined as 
\begin{equation}
    \bar{\tau}_t^{\overrightarrow{k},l}  = min(\tau_t^{\overrightarrow{k},l} , K_{l-1}-1).
\end{equation}
When replacing $\tau_t^{\overleftarrow{k},l}$ with $\bar{\tau}_t^{\overleftarrow{k},l}$ in \eqref{eq:for_for_for_proof} and \eqref{eq:back_for_for_proof}, we get
\begin{equation}
    \label{eq:bi_connection_for_diffK}
    \overrightarrow{k} \leftarrow \left(
                                    \begin{array}{c}
                                        (t- \bar{\tau}_t^{\overrightarrow{k},l})\ \text{mod}\ K_{l-1} \\ 
                                        ((T-1)-t-\bar{\tau}_t^{\overrightarrow{k},l})\ \text{mod}\ K_{l-1}
                                    \end{array}
                                \right).
\end{equation}
Similarly, for the backward direction we define
\begin{equation}
    \bar{\tau}_t^{\overleftarrow{k},l} = min(\tau_t^{\overleftarrow{k},l},K_{l-1}-1),
\end{equation}
\begin{equation}
    \label{eq:bi_connection_back_diffK}
    \overleftarrow{k} \leftarrow \left(
                                    \begin{array}{c}
                                        (t-\bar{\tau}_t^{\overleftarrow{k},l})\ \text{mod}\ K_{l-1} \\ 
                                        ((T-1)-t-\bar{\tau}_t^{\overleftarrow{k},l})\ \text{mod}\ K_{l-1}  
                                    \end{array}
                                \right).
\end{equation}
With the constraint $K_l \geq K_{l-1}$ (otherwise, layer $l$ would be allowed less context than layer $l-1$. Instances with $\tau_t^{k,l-1}>K_l-1$ would not be connected to the higher layer and effectively $K_{l-1}$ would be set to $K_{l}$).

\section{Grouped inter-layer connections}
\label{app:bewijs_grouped}
Using \eqref{eq:kstar_group}, instance $k$ of layer $l$ will be reset at times $t_{k,l}$ given by
\begin{equation}
\label{eq:t_for_k_groups}
    t_{k,l} = (k + \alpha K_l)G_l.
\end{equation}
Therefore, the number of time steps $\tau_t^{k,l}$ between time $t$ and the last time instance $k$ was reset before time $t$ is given by
\begin{equation}
\begin{split}
    \label{eq:tau_def_proof_groups}
    \tau_t^{k,l} &= (t-t_{k,l})\ \text{mod}\ T_{\text{reset}}^l \\
    &= (t-kG_l-\alpha K_lG_l)\ \text{mod}\ K_lG_l \\
    &= (t-kG_l)\ \text{mod}\ K_lG_l.
\end{split}
\end{equation} 
As before, we would like an instance to receive input from an instance in the layer below, with the same number of context frames. However, this cannot be guaranteed as $\tau_t^{k,l}$ increases with steps of $G$. For the forward input we therefore select the first instance in layer $l-1$ to be reset at $t-\tau_t^{k,l}$ or afterwards. \eqref{eq:t_for_k_groups} shows that resets happen at multiples of $G_{l-1}$. Therefore the requested reset in the forward direction will happen at time
\begin{equation}
\label{eq:gamma_group}
    \overrightarrow{\gamma}_t^{\overrightarrow{k},l}=\ceil*{\frac{t-\bar{\tau}_t^{\overrightarrow{k},l}}{G_{l-1}}}G_{l-1},
\end{equation}
where $\bar{\tau}$ is defined as
\begin{equation}
    \bar{\tau}_t^{\overrightarrow{k},l}  = min(\tau_t^{\overrightarrow{k},l} , K_{l-1}G_{l-1}-1).
\end{equation}
Using \eqref{eq:kstar_group}, we find that the instance in the forward direction in layer $l-1$ that will be reset at time $\overrightarrow{\gamma}_t^{\overrightarrow{k},l}$ is given by
\begin{equation}
\begin{split}
     \overrightarrow{k}^{*,l-1}_{\overrightarrow{\gamma}_t^{\overrightarrow{k},l}} &= \frac{\overrightarrow{\gamma}_t^{\overrightarrow{k},l}}{G_{l-1}}\ \text{mod}\ K_{l-1} \\
     &=\left(\ceil*{\frac{t-\bar{\tau}_t^{\overrightarrow{k},l}}{G_{l-1}}} \frac{G_{l-1}}{G_{l-1}}\right)\ \text{mod}\ K_{l-1} \\ &=\ceil*{\frac{t-\bar{\tau}_t^{\overrightarrow{k},l}}{G_{l-1}}}\ \text{mod}\ K_{l-1}.
\end{split}
\end{equation}
In the backward direction \eqref{eq:kstar_group} is changed to
\begin{multline}
\label{eq:kstar_group_back}
    \overleftarrow{k}^{*,l}_t=\left(\frac{(T-1)-t}{G_l}\right)\ \text{mod} \ K_l \ \\ \text{if} \ (T-1)-t \equiv 0\ (\text{mod}\ G_l).
\end{multline}
The requested reset in the backward direction will happen at time
\begin{equation}
\label{eq:gamma_group_back}
    \overleftarrow{\gamma}_t^{\overrightarrow{k},l}=(T-1)-\left(\ceil*{\frac{(T-1)-(t+\bar{\tau}_t^{\overrightarrow{k},l})}{G_{l-1}}}G_{l-1}\right).
\end{equation}
Combining \eqref{eq:kstar_group_back} and \eqref{eq:gamma_group_back} gives the instance in the backward direction in layer $l-1$ that will be reset at time $\overleftarrow{\gamma}_t^{\overrightarrow{k},l}$.
\begin{equation}
\begin{split}
\overleftarrow{k}^{*,l-1}_{\overleftarrow{\gamma}_t^{\overrightarrow{k},l}} &= \left(\frac{(T-1)-\overleftarrow{\gamma}_t^{\overrightarrow{k},l}}{G_{l-1}}\right)\ \text{mod} \ K_{l-1}  \\
&=  \ceil*{\frac{(T-1)-(t+\bar{\tau}_t^{\overrightarrow{k},l})}{G_{l-1}}}\ \text{mod} \ K_{l-1}.
\end{split}
\end{equation}
In shorthand notation this becomes
\begin{equation}
    \label{eq:bi_connection_for_diffK_groups}
    \overrightarrow{k} \leftarrow \left(
                                    \begin{array}{c}
                                        \ceil*{\frac{t-\bar{\tau}_t^{\overrightarrow{k},l}} {G_{l-1}}}\ \text{mod}\ K_{l-1} \\ 
                                        \ceil*{\frac{(T-1)-t-\bar{\tau}_t^{\overrightarrow{k},l}}{G_{l-1}}}\ \text{mod}\ K_{l-1}
                                    \end{array}
                                \right).
\end{equation}
Similarly, for the backward direction we find the inputs to be
\begin{equation}
    \label{eq:bi_connection_back_diffK_groups}
    \overleftarrow{k} \leftarrow \left(
                                    \begin{array}{c}
                                        \ceil*{\frac{t-\bar{\tau}_t^{\overleftarrow{k},l}}{ G_{l-1}}}\ \text{mod}\ K_{l-1} \\ 
                                        \ceil*{\frac{(T-1)-t-\bar{\tau}_t^{\overleftarrow{k},l}}{G_{l-1}}}\ \text{mod}\ K_{l-1}
                                    \end{array}
                                \right).
\end{equation}

Finally, we would like to find the final output of the network or a generalization of \eqref{eq:reset_BLSTM_out_proof}. Ideally, we would like to select the instances with $T_{reset,L}-1=K_{L}G_{L}-1$ number of context frames. Again this cannot be guaranteed as $\tau_t^{\overrightarrow{k},L}$ and $\tau_t^{\overleftarrow{k},L}$ increase with steps of $G_L$. Instead we will be looking for time $\overrightarrow{\gamma}_t^{\overrightarrow{max},L}$ and $\overleftarrow{\gamma}_t^{\overleftarrow{max},L}$ defined as

\begin{equation}
\label{eq:gamma_group_max_L_for}
    \overrightarrow{\gamma}_t^{\overrightarrow{max},L}=\ceil*{\frac{t-(K_{L}G_{L}-1)}{G_{L}}}G_{L}
\end{equation}
and
\begin{multline}
\label{eq:gamma_group_max_L_back}
    \overleftarrow{\gamma}_t^{\overleftarrow{max},L}=(T-1)- \\ \left(\ceil*{\frac{(T-1)-(t+(K_{L}G_{L}-1))}{G_{L}}}G_{L}\right).
\end{multline}

The corresponding instances are thus
\begin{equation}
\label{eq:reset_BLSTM_diffK_groups_out_for_proof}
\begin{split}
\overrightarrow{k}^{*,L}_{\overrightarrow{\gamma}_t^{\overrightarrow{max},L}} &= \frac{\overrightarrow{\gamma}_t^{\overrightarrow{max},L}}{G_{L}}\ \text{mod}\ K_{L} \\
     &=\left(\ceil*{\frac{t-(K_{L}G_{L}-1)}{G_{L}}} \frac{G_{L}}{G_{L}}\right)\ \text{mod}\ K_{L} \\ 
     &=\ceil*{\frac{t-(K_{L}G_{L}-1)}{G_{L}}}\ \text{mod}\ K_{L}.
\end{split}
\end{equation}
and
\begin{equation}
\label{eq:reset_BLSTM_diffK_groups_out_back_proof}
\begin{split}
\overleftarrow{k}^{*,L}_{\overleftarrow{\gamma}_t^{\overleftarrow{max},L}} &= \left(\frac{(T-1)-\overleftarrow{\gamma}_t^{\overleftarrow{max},L}}{G_{L}}\right)\ \text{mod} \ K_{L}  \\
&=  \ceil*{\frac{(T-1)-(t+(K_{L}G_{L}-1))}{G_{L}}}\ \text{mod} \ K_{L}.
\end{split}
\end{equation}

\section*{Funding}
This research was funded by Research Foundation Flanders with grant number 1S66217N.


\bibliographystyle{abbrvnat}
\setcitestyle{authoryear}
\bibliography{mybib.bib}




\end{document}